\documentclass[twocolumn,aps,prd,preprintnumbers,showpacs,superscriptaddress,nofootinbib,amsmath,amssymb,floats,floatfix,showkeys,notitlepage,longbibliography]{revtex4-1}
\usepackage{comment}
\usepackage{graphicx}
\usepackage{subfigure}
\usepackage{palatino}
\usepackage[commandnameprefix=always]{changes}
\usepackage{hyperref}
\hypersetup{colorlinks=true,linkcolor=blue,urlcolor=blue,citecolor=blue}
\usepackage[toc,page]{appendix}
\usepackage[normalem]{ulem}

\usepackage{orcidlink}
\usepackage{lipsum}
\usepackage{graphicx}
\usepackage{subfigure}
\usepackage{palatino}
\usepackage{sans}
\usepackage{adjustbox}
\usepackage{latexsym}
\usepackage{amsmath}
\usepackage{amssymb}
\usepackage{amsfonts}
\usepackage{dcolumn}
\usepackage{bm}
\usepackage{tikz}
\usepackage{bigints}
\usepackage{array,tabularx,multirow,booktabs}
\usepackage[tracking=true]{microtype}
\SetTracking{}{500}
\SetTracking{encoding={*}, shape=sc}{40}
\UseRawInputEncoding 
\allowdisplaybreaks
\usepackage{adjustbox}
\usepackage{latexsym}
\usepackage{amsmath}
\usepackage{amssymb}
\usepackage{amsfonts}
\usepackage{dcolumn}
\usepackage{bm}
\usepackage{tikz}
\usepackage{bigints}
\usepackage{array,tabularx,multirow,booktabs}
\usepackage[tracking=true]{microtype}
\usepackage{color}
\UseRawInputEncoding 
\allowdisplaybreaks

\begin{document} \sloppy

\title{The Effect of Quark-antiquark Confinement on the Deflection Angle by the NED Black Hole}

\author{Erdem Sucu
\orcidlink{0009-0000-3619-1492}
}
\email{erdemsc07@gmail.com}
\affiliation{Physics Department, Eastern Mediterranean
University, Famagusta, 99628 North Cyprus, via Mersin 10, Turkiye}

\author{Ali \"Ovg\"un
\orcidlink{0000-0002-9889-342X}
}
\email{ali.ovgun@emu.edu.tr}
\affiliation{Physics Department, Eastern Mediterranean
University, Famagusta, 99628 North Cyprus, via Mersin 10, Turkiye}

\begin{abstract}

In this study, we explore the influence of quark-antiquark confinement on the deflection angle within the framework of nonlinear electrodynamic (NED) black holes. To achieve this, we establish the appropriate optical spacetime metric and subsequently determine the Gaussian optical curvature. Utilizing the Gauss-Bonnet theorem, we investigate the impact of quark-antiquark confinement on the deflection angle exhibited by NED black holes. Additionally, we delve into the effects of a cold non-magnetized plasma medium and also axion-plasmon on gravitational lensing. Our findings highlight the significance of the axion-plasmon effect on the optical properties of NED black holes, particularly its influence on gravitational lensing. This exploration is particularly relevant in the context of the axion's potential role as a dark matter candidate. The multifaceted interplay between quark-antiquark confinement, nonlinear electrodynamics, and plasma dynamics provides a nuanced understanding of gravitational lensing phenomena. These insights contribute to ongoing research in dark matter studies and offer avenues for further theoretical and observational investigations in astrophysics.

\end{abstract}

\date{\today}

\keywords{Black hole; Confinement;
Nonlinear electrodynamics;
Dark matter; Deflection angle; Quark; Axion.}

\pacs{95.30.Sf, 04.70.-s, 97.60.Lf, 04.50.Kd }

\maketitle

\section{Introduction}

Despite its numerous successes, General Relativity (GR) faces limitations, particularly within the framework of standard black hole (BH) solutions \cite{LIGOScientific:2016aoc,EventHorizonTelescope:2019dse}. GR predicts the existence of singularities, regions where the laws of physics break down, casting doubt on the validity of Einstein's theory \cite{Kerr:2023rpn,Penrose:1964wq,Hawking:1971vc}. To address these shortcomings, researchers have explored the possibility of singularity-free BH solutions within GR by considering alternative matter distributions.  Nonlinear electrodynamics (NED) models offer potential extensions of linear electrodynamics, particularly in the high-energy regime where electromagnetic fields approach strong intensities. One of the first NED models, Born-Infeld electrodynamics, emerged in 1934 as a means to obtain a finite self-energy density for the electric charge \cite{Born:1934ji,Born:1934gh}. Another prominent NED model, Euler-Heisenberg theory, is linked to two significant aspects of quantum electrodynamics (QED): light-by-light scattering and vacuum birefringence \cite{Heisenberg:1936nmg,Mignani:2016fwz}. NED's applications extend beyond black hole physics and QED, encompassing string/M-theories \cite{Fradkin:1985qd,Seiberg:1999vs,Tseytlin:1999dj,Gibbons:2000xe} and cosmology \cite{Novello:2003kh,Garcia-Salcedo:2000ujn,Novello:2006ng,Kruglov:2015fbl,Ovgun:2017iwg,Ovgun:2016oit,Otalora:2018bso,Benaoum:2021tec,Benaoum:2022uta}. 

In 1968, James Bardeen proposed the first line element for a non-singular BH geometry. This work paved the way for the development of various exact charged regular black hole (RBH) solutions, achieved through minimal coupling of GR with nonlinear electrodynamics (NED) \cite{Hayward:2005gi,Ayon-Beato:1999qin,Dymnikova:1992ux,Frolov:2016pav,Bronnikov:2000vy,Bronnikov:2005gm,Ma:2015gpa,Nascimento:2019qor}.  
 Within these theories, the Bardeen geometry can be interpreted as an RBH sourced by a nonlinear magnetic or electric monopole. Nonlinear electrodynamics (NED) \cite{javed2023effects} models have been thoroughly examined as potential frameworks (such as Cornell potential \cite{eichten1978charmonium}) to uncover novel aspects of physics. 

One of the key applications of Einstein's theory of general relativity (GR) \cite{einstein2013principle} is gravitational lensing, a phenomenon in which the path of a light beam is bent by the presence of a massive object, and this object acting as the bending agent is referred to as a gravitational lens. Gravitational lensing \cite{jusufi2018gravitational,bartelmann2010gravitational,jusufi2019gravitational,Mangut:2023oxa,Atamurotov:2022knb,Abdujabbarov:2017pfw,Atamurotov:2021imh,Atamurotov:2021hoq,Atamurotov:2021qds,Vagnozzi:2022moj,Vagnozzi:2019apd,Allahyari:2019jqz,Khodadi:2020jij,Virbhadra:1999nm,Virbhadra:2002ju,Virbhadra:2008ws,Nascimento:2020ime,Furtado:2020puz} is a helpful technique to understand black holes, galaxies, dark matter, dark energy, and the universe. Additionally, research on gravitational lensing systems plays a crucial role in understanding the cosmic microwave background radiation and its associated cosmological elements, as highlighted in references \cite{schneider2006introduction,hanson2010weak}.

\textcolor{black}{Gravitational lensing in a strong gravitational field} \cite{Virbhadra:1998dy,Virbhadra:1999nm,Virbhadra:2002ju,2006glsw.conf.....M,bozza2002gravitational,kuang2022strong} allows us to identify the location, magnification, and time delays of the images produced by black holes. Additionally, the effect is somewhat smaller with gravitational lensing \cite{bartelmann2001weak,ovgun2019weak}, although it is still statistically significant. \textcolor{black}{ The consideration of deflection angles holds significance not only in a vacuum but also within a plasma medium. Bisnovatyi-Kogan and Tsupko conducted an in-depth exploration of gravitational lensing within a plasma medium. \cite{Bisnovatyi-Kogan:2010flt}.}

The most utilised technique for determining the deflection angle is the Gauss-Bonnet theorem (GBT), which was first presented in ref \cite{gibbons2008applications}. After that, the GBT has been shown to be quite helpful in determining the deflection angle of different types of black holes that exhibit asymptotic behaviour \cite{ovgun2018gravitational,javed2022weak,javed2021weak,kumaran2021deriving,ovgun2022effect,Lambiase:2023zeo,Ovgun:2023wmc,Pulice:2023dqw,Lambiase:2023hng,Ovgun:2023ego,Pantig:2022gih,Pantig:2022qak,Rayimbaev:2022hca,Uniyal:2022vdu}.

While it is widely accepted that astrophysical black holes (BHs) are essentially neutral, some researchers have suggested the possibility of a small non-zero electric charge for these objects. This charge can influence the motion of charged particles, making the study of electrically charged BHs, particularly in the spherically symmetric case, crucial for advancing our understanding of BH physics. Moreover, such investigations can assess the role of nonlinear electrodynamics (NED) and its potential implications for astrophysical BHs. 

This work aims to explore the imprints of different mediums, including cold non-magnetized, axion plasmon, and homogenous magnetized, on the trajectories of photons as analyzed through gravitational lensing. 
Often referred to as weakly interacting scalar particles, axions are the lightest and coldest particles that interact extremely weakly with photons and the conventional model of particles. It is also commonly known that in the presence of a magnetic field, axions interact with the plasma as well.

For the purpose of this study, we focus on the static and spherically symmetric electrically charged NED black hole with quark-antiquark confinement spacetime, a specific model that incorporates elements of both gravity and electromagnetism.

The paper is organized as follows: Section (2) introduces the NED black hole with quark-antiquark confinement spacetime and analyzed its physical features. In section (3) we have derived the effect of small gravitational lensing caused by a NED Black Hole. This analysis pertains to situations where the observer and the light source are situated in a region called asymptotic safety. In section (4)  it is reviewed the Jacobi metric for Quark-antiquark particles in NED spacetime and then find a small  deflection angle. In section (5) we calculate the deflection angle of the NED  black hole by utilizing the Gauss-Bonnet theorem in a cold non-magnetized plasma medium. In section (6) we consider the axion plasmon effect on the NED BH in the presence of a homogenous magnetized, and we proceed to calculate the small  deflection angle by using GBT. Finally, we draw our conclusions in section(7).

\section{NED black hole with quark-antiquark confinement} \label{sec7}

The Nonlinear electrodynamics (NED) model of Einstein's gravity within the following action can be expressed as follows\cite{Mazharimousavi:2023okd}

 \begin{eqnarray}
 S=\int d^4 x\sqrt{-g}\left[\frac{\mathcal{R}}{16\pi G} + \mathcal{L} \right], \label{izm1}
 \end{eqnarray}

where $\mathcal{R}$ stands for the Ricci scalar, G=1 and   $\mathcal{L}$ is the NED Lagrangian with Quark-antiquark confinement

\begin{equation}
\mathcal{L}=-\frac{16\left(3 \sqrt{-2 \mathcal{F}}+\zeta\left(\zeta+\sqrt{\zeta^{2}+4 \sqrt{-2 \mathcal{F}}}\right)\right) \sqrt{-2 \mathcal{F}}}{3\left(\zeta+\sqrt{\zeta^{2}+4 \sqrt{-2 \mathcal{F}}}\right)^{4}} \mathcal{F}
\end{equation}
 with $
\mathcal{F} \equiv \frac{1}{4} F_{\mu \nu} F^{\mu \nu}
=-\frac{1}{2}\left(\frac{q}{r^{2}}+\frac{f}{r}\right)^{2}$ and $
f=\zeta \sqrt{q} \text { with } 
\zeta
$ is a new constant related with $f$ quark-antiquark confinement constant.
The Einstein field equation can be written as

\begin{equation}
G_{\mu}^{\nu}=8 \pi T_{\mu}^{\nu}
\end{equation} with the energy momentum tensor of the nonlinear electromagnetic field with Quark-antiquark confinement

\begin{equation}
T_{\mu}^{v}=\frac{1}{4 \pi}\left(\mathcal{L} \delta_{\mu}^{v}-\mathcal{L}_{\mathcal{F}} F_{\mu \lambda} F^{v \lambda}\right)
\end{equation} where $
\mathcal{L}_{\mathcal{F}}=\frac{\partial \mathcal{L}}{\partial \mathcal{F}}
$.

We are looking for a static and spherically symmetric space-time geometry whose line element is given by

\begin{equation}
ds^{2}=-A(r)dt^{2}+(A(r))^{-1}dr^{2}+r^{2}d\theta ^{2}+r^{2}sin^{2}(\theta)d\phi ^{2}
\label{}
\end{equation}
in which 
\begin{equation}
A(r)=1-\frac{2 M}{r}+\frac{q^{2}}{r^{2}}-\frac{4 q \sqrt{q} \zeta}{3 r} \ln (r) \label{metric}
\end{equation}

The electric charge of the black hole or particle is denoted by $q$, while $M$ represents an integration constant linked to its mass. The nature of the spacetime, whether asymptotically flat or exhibiting a singular black hole or a naked singularity, depends on the values of the NED parameter $\zeta$, mass, and charge. \textcolor{black}{Note that for Reissner - Nordsrom metric it has been studied analytical expression for shadow size as a function of charge in \cite{Zakharov:2014lqa,Zakharov:2005ek}.} 


The black hole's physical properties are believed to resemble those of a Reissner-Nordstrom black hole with a correction term proportional to $\zeta$.The black hole exhibits: (1) two distinct horizons (interior and exterior) for  M $>$ $M_c$, (2) a single degenerate horizon for M = $M_c$, and (3) the absence of any horizon for M $<$  $M_c$.

\begin{equation}
M_{c}=(1-ln\Delta)\frac{\zeta q\sqrt{q}}{3}+   \frac{q^{2}}{\Delta}
\label{}
\end{equation}
where

\begin{equation}
\Delta= (1+ \sqrt{1+\frac{9}{4q\zeta^{2}}}) \frac{2\zeta q\sqrt{q}}{3}
\label{}
\end{equation}

\subsection{ Topological approach to derive the global Hawking temperature of NED black hole with quark-antiquark confinement}

Using the topological technique, one can determine the Hawking temperature without sacrificing any knowledge about the higher-dimensional space by using the Euclidean geometry of the 2-dimensional spacetime. The thermodynamic property of Hawking temperature for a two-dimensional black hole can be established using the topological method \cite{Robson:2018con,Ovgun:2019ygw,Zhang:2020kaq}
\begin{equation}
T_{\text{H}}=\frac{\hbar c}{4\pi \chi k_{\text{B}}}\Sigma_{j\leq\chi}\int_{r_{h_j}}{\sqrt{|g|}\mathcal{R}dr}.
\label{temperature}
\end{equation}
In this context, the symbols $\hbar$, $c$, and $k_{\text{B}}$ represent the Planck constant, speed of light, and Boltzmann's constant, respectively. Additionally, $g$ corresponds to the metric determinant, and $r_{h_j}$ signifies the $j$-th killing horizon. For the purpose of this study, we adopt the values $\hbar=1$, $c=1$, and $k_{\text{B}}=1$ for these parameters. The function $\mathcal{R}$ denotes the Ricci scalar in the two-dimensional spacetime. The variable $\chi$ represents the Euler characteristic of the Euclidean geometry and is linked to the count of Killing horizons. The symbol $\Sigma_{j\leq\chi}$ signifies the summation across the Killing horizons.

The Euler characteristic in a two-dimensional spacetime is expressed as follows:

\begin{equation}
\chi=\int \sqrt{|g|} d^2x\frac{\mathcal{R}}{4\pi}.
\end{equation}

Upon employing the Wick rotation $t=i\tau$ and defining the new compact time as the inverse temperature $\beta$, the Euler characteristic $\chi$ undergoes a transformation \cite{Robson:2018con,Ovgun:2019ygw}.
\begin{equation}
\chi=\int_0^\beta{d\tau}\int_{r_{\text{H}}}\sqrt{|g|}dr \frac{\mathcal{R}}{4\pi}.
\end{equation}

Subsequently, the connection between the Hawking temperature $T_{H}$ and the Euler characteristic $\chi$ is established through the relation:

\begin{equation}
\frac{1}{4\pi T_{H}}\int_{r_{\text{H}}}{\sqrt{|g|}\mathcal{R}dr}=\chi,
\end{equation}

this relationship serves as the basis for Eq. (\ref{temperature}). 

By considering a specific hypersurface, the black hole can be transformed into a two-dimensional configuration with a reduced metric \cite{Achucarro:1993fd} through the Wick rotation ($\tau = i t$):
\begin{equation}
ds^2=A(r)d\tau^2+\frac{dr^2}{A(r)}.
\label{weakmetric}
\end{equation}

The Ricci scalar corresponding to the reduced metric (\ref{weakmetric}) is given by:
\begin{equation}
\mathcal{R}=-\frac{d^2}{dr^2}A(r)= \frac{4q^{3/2} \zeta}{3r}+ \frac{4M}{r^{3}} - \frac{6q^{2}}{r^{4}}.
\end{equation}

Hence, the temperature of the black hole is determined by employing the formula:
\begin{equation}
T_{H}=\frac{1}{4 \pi \chi} \int_{r_{+}} \sqrt{|g|} R d r=\frac{1}{4 \pi r_{+}}-\frac{q^{2}}{4 \pi r_{+}^{3}}-\zeta \frac{q^{3 / 2}}{3 \pi r_{+}^{2}}
\end{equation}

\section{Gauss-Bonnet theorem and optical metric to calculate the deflection angle of NED black hole with quark-antiquark confinement}

In this section, we undertake a review of the Gauss-Bonnet theorem and proceed to compute the small deflection angle for the Nonlinear Electrodynamics (NED) black hole incorporating quark-antiquark confinement. Initially, we express the null geodesics satisfying $ds^2=0$, a rearrangement of which yields:

\begin{eqnarray}
dt^2=\gamma_{ij}dx^i dx^j=\frac{1}{A^2}dr^2+\frac{r^2}{A}d\Omega^2,~\label{opmetric}
\end{eqnarray}

Here, $i$ and $j$ range from $1$ to $3$, and $\gamma_{ij}$ represents the optical metric. Following a coordinate transformation $dr^*=\frac{1}{A}dr$, the aforementioned expression can be reformulated as:

\begin{eqnarray}
dt^2=dr^{*2}+\tilde{A}^2(r^*)d\phi^2,
\end{eqnarray}
where $\tilde{A}(r^*)\equiv\sqrt{\frac{r^2}{A}}$ and $\theta=\frac{\pi}{2}$.
\begin{widetext}

To employ the Gauss-Bonnet theorem, it is imperative to compute the Gaussian curvature, and this calculation is performed here:
\begin{eqnarray}
\mathcal{K}&=&\frac{R_{r\phi r\phi}}{\gamma}=\frac{1}{\sqrt{\gamma}}\left[\frac{\partial}{\partial \phi}\left(\frac{\sqrt{\gamma}}{\gamma_{rr}}\Gamma^{\phi}_{rr}\right)-\frac{\partial}{\partial r}\left(\frac{\sqrt{\gamma}}{\gamma_{rr}}\Gamma^{\phi}_{r\phi}\right)\right] \\
\mathcal{K}=\frac{R}{2}&=& \frac{A\! \left(r\right)}{2} \frac{d^{2}}{d r^{2}}A\! \left(r\right) -\frac{\left(\frac{d}{d r}A\! \left(r\right)\right)^{2}}{4}=\frac{3 M^2}{r^4}-\frac{8 \zeta  M q^{3/2}}{3 r^4}+\frac{4 \zeta  M q^{3/2} \log (r)}{r^4}-\frac{6 M q^2}{r^5}-\frac{2 M}{r^3}+\frac{2 \zeta  q^{7/2}}{3 r^5}-\frac{4 \zeta  q^{7/2} \log (r)}{r^5}\notag\\&+&\frac{2 \zeta  q^{3/2}}{r^3}-\frac{4 \zeta  q^{3/2} \log (r)}{3 r^3}+\frac{2 q^4}{r^6}-\frac{4 \zeta ^2 q^3}{9 r^4}+\frac{4 \zeta ^2 q^3 \log ^2(r)}{3 r^4}-\frac{16 \zeta ^2 q^3 \log (r)}{9 r^4}+\frac{3 q^2}{r^4}.~\label{GC}
\end{eqnarray}
Here, $\gamma\equiv\det (\gamma_{ij})$, and $R$ represents the Ricci scalar.
\end{widetext}

\begin{figure}
    \centering
    \includegraphics[width=0.5\linewidth]{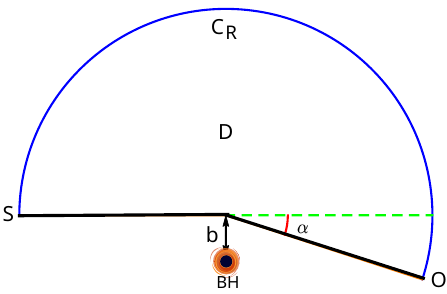}
    \caption{ In the depicted diagram, the source is denoted by the point S, and the observer corresponds to O. The black line represents a light ray emitted by the source and reaching the observer at O. The parameter "b" is associated with the impact parameter.}
    \label{fig:enter-label}
\end{figure}

Consider the domain $D$ as a compact, oriented, nonsingular two-dimensional Riemannian surface with Euler characteristic $\chi(D)$ and Gaussian curvature $\mathcal{K}$. This domain is bounded by a piecewise smooth curve with geodesic curvature $\kappa$. The connection between the deflection angle of light and the Gaussian curvature is established through the Gauss-Bonnet theorem, which is applied by utilizing:

\begin{eqnarray}
\int\int_D \mathcal{K}dS+\oint_{\partial D}\kappa dt+\sum_{i=1}\beta_i=2\pi \chi(D),~\label{GB}
\end{eqnarray}

Here, $dS$ represents the surface element, $\kappa$ stands for the geodesic curvature of the boundary, defined as $\kappa=|\nabla_{\dot{C}}\dot{C}|$, and $\beta_i$ denotes the $i^{\text{th}}$ exterior angle. For a specific region $\tilde{D}$ bounded by a geodesic $C_1$ from the source $S$ to the observer $O$ and a circular curve $C_R$ intersecting $C_1$ at right angles at $S$ and $O$, Equation (\ref{GB}) simplifies to:
\begin{eqnarray}
\int\int_{\tilde{D}}\mathcal{K}dS+\int_{C_R}\kappa(C_R)dt=\pi,~\label{GB2}
\end{eqnarray}

In this derivation, we utilized $\kappa(C_1)=0$ and the Euler characteristic $\chi(\tilde{D})=1$. Specifically, for the circular curve $C_R := r(\phi)=R=\text{const}$, the non-zero segment of the geodesic curvature can be computed as:

\begin{eqnarray}
\kappa(C_R)=\left(\nabla_{\dot{C}_R}\dot{C}_R\right)^r=\dot{C}^{\phi}_R(\partial_{\phi}\dot{C}^r_R)+\Gamma^r_{\phi\phi}(\dot{C}^{\phi}_R)^2,
\end{eqnarray}

Here, $\dot{C}_R$ represents the tangent vector of the circular curve $C_R$, and $\Gamma^r_{\phi\phi}$ is the Christoffel symbol associated with the optical metric (\ref{opmetric}). In the final equation, it is evident that the first term vanishes, and $\Gamma^r_{\phi\phi}=-\tilde{A}(r^*)\tilde{A}'(r^*)$, with $(\dot{C}^{\phi}_R)^2=\frac{1}{\tilde{A}^2(r^*)}$ in the second term. As $R$ approaches infinity, one obtains:

\begin{eqnarray}
&&\lim_{R\rightarrow \infty}\left[\kappa(C_R)dt\right]=\lim_{R\rightarrow \infty}[-\tilde{A}'(r^*)]d\phi=d\phi.~\label{geoR}
\end{eqnarray}

Substituting Eq.~(\ref{geoR}) into Eq.~(\ref{GB2}), we obtain:
\begin{eqnarray}
\int\int_{\tilde{D}_{R\rightarrow \infty}}\mathcal{K}dS+\int_0^{\pi+\alpha}d\phi=\pi.
\end{eqnarray}

Here, the surface area on the equatorial plane is expressed as \cite{Gibbons:2008rj,Molla:2023hou}:

\begin{equation}
dS=\sqrt\gamma dr d \phi= \frac{r}{A^{3/2}} dr d\phi
\label{}
\end{equation}

\begin{figure}
    \centering
\includegraphics[width=0.4\textwidth]{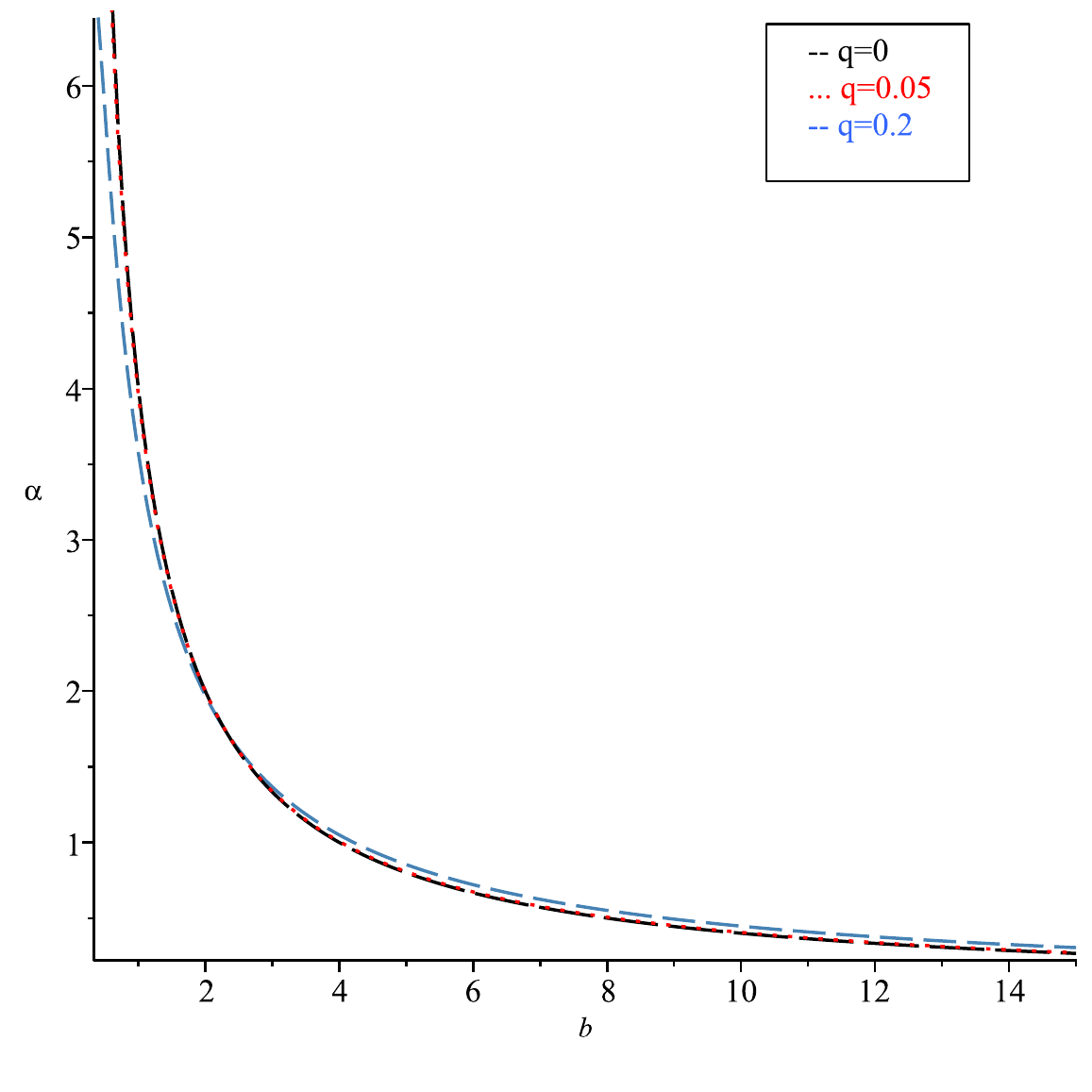}
 \caption{In the figure, the deflection angle $\alpha$ in the weak gravitational field limit is depicted as a function of $b$ for the case of $\zeta=1$, $M=1$, and $q=0.2$.}
    \label{section3}
\end{figure}

\begin{widetext}

Subsequently, the deflection angle of light can be computed as:

\begin{eqnarray}
\alpha&=&-\int\int_{\tilde{D}}\mathcal{K}dS=-\int^{\pi}_0\int^{\infty}_{\frac{b}{\sin\phi}}\mathcal{K}dS \nonumber\\
&\simeq&\frac{4 M}{b}-\frac{4 M q^2}{3 b^3}-\frac{5 \pi  \zeta  M q^{3/2}}{6 b^2}-\frac{3 \pi  q^2}{4 b^2}+\frac{4 \zeta  q^{3/2}}{3 b}+\frac{8 \zeta  q^{3/2} \log (b)}{3 b}-\frac{4 \zeta  q^{3/2} \log (4)}{3 b}+\mathcal{O}(M^2,q^2).~\label{deflang}
\end{eqnarray}

In this calculation, the zero-order particle trajectory $r=b/\sin\phi$, where $0\leq\phi\leq\pi$ in the weak deflection limit, has been employed which is shown in Fig.\ref{section3}.

\end{widetext}

\section{Jacobi geometry to calculate deflection angle of NED black hole with quark-antiquark confinement by massive particles}

\textcolor{black}{Following the principle of least action proposed by Maupertuis, Gibbons \cite{Gibbons:2015qja}, Chanda, Gibbons and Guha ~\cite{Chanda:2016aph}} used the Jacobi metric framework for curved spacetime. The motion of null massive particles in a background spacetime can be represented as a spatial geodesic in the corresponding Jacobi geometry defined by the Jacobi metric. This parallels the scenario in which the motion of photons can be described as a spatial geodesic in the corresponding optical geometry \cite{Crisnejo:2018uyn,Li:2019vhp}. Even in the case of charged particles~\cite{Das:2016opi}, the Jacobi metric approach remains applicable. Hence, the Jacobi geometry can be employed as a background space for studying the deflection of particles.

For a static metric
\begin{eqnarray}
d\bar s^2=\bar g_{tt}dt^2+\bar g_{ij}dx^i dx^j,
\end{eqnarray}
the corresponding Jacobi metric reads~\cite{Gibbons:2015qja}
\begin{eqnarray}
\label{sjmetric}
&& g_{ij}=\left(E^2+m^2 \bar g_{tt}\right) g^{\mathrm{opt}}_{ij}.
\end{eqnarray}

Here, $E$ and $m$ denote the particle energy and mass, respectively, and $g^{\mathrm{opt}}_{ij}$ represents the corresponding optical metric of the static metric as provided by~\cite{Gibbons:2008rj,Crisnejo:2018uyn}:

\begin{eqnarray}
&& g^{opt}_{ij}=-\frac{\bar g_{ij}}{\bar g_{tt}}.
\end{eqnarray}

It's worth noting that the Jacobi metric in Eq.~\eqref{sjmetric} is essentially identical to a specific optical metric associated with massive particles presented in Ref.~\cite{Gibbons:2008rj}.

The general expression for a static and spherically symmetric metric is given by:
\begin{equation}
\label{SS-metric}
d\bar s^2=-A\left(r\right)dt^2+\frac{1}{A\left(r\right)}dr^2+C\left(r\right)d\Omega^2.\\
\end{equation}

Here, $d\Omega^2=d\theta^2+\sin^2\theta d\varphi^2$ represents the line element of the unit two-sphere. According to Eq.~\eqref{sjmetric}, its corresponding Jacobi metric is:

\begin{eqnarray}
\label{SS-Jacobi-1}
ds^2&=&\bigg(E^2-m^2A\bigg)\bigg[\frac{1}{A^2}dr^2+\frac{C}{A}d\Omega^2 \bigg].
\end{eqnarray}

Exploiting spherical symmetry, we focus solely on the motion of massive particles in the equatorial plane $\theta=\pi/2$ without loss of generality. Consequently, the Jacobi metric is then given by:

\begin{equation}
\label{SS-Jacobi-2}
ds^2=\bigg(E^2-m^2A\bigg)\bigg(\frac{1}{A^2}dr^2+\frac{C}{A}d\varphi^2 \bigg).\\
\end{equation}

The conserved angular momentum $J$ can be derived by leveraging axial symmetry:

\begin{equation}
\label{angular momentum}
J=\left(E^2-m^2A\right)\frac{C}{A}\left(\frac{d\varphi}{ds}\right)= \mathrm{constant},
\end{equation}
together with Eq.~\eqref{angular momentum} and Eq.~\eqref{SS-Jacobi-2}, which yields
\begin{equation}
\label{radial equation}
\left(E^2-m^2A\right)^2\frac{1}{A^2}\left(\frac{dr}{ds}\right)^2=E^2-A\left(m^2+\frac{J^2}{C}\right).
\end{equation}

Then one finds
\begin{equation}
m^2\left(\frac{dr}{d\tau}\right)^2=E^2-A\left(m^2+\frac{J^2}{C}\right).
\end{equation}

Here, $\tau$ is employed to denote the proper time along the geodesic, and subsequently:
\begin{eqnarray}
E=m A\frac{dt}{d\tau}\label{conserved quantity 1}~,\ \ J=m C\frac{d\varphi}{d\tau},
\end{eqnarray}
with
\begin{equation}
\label{proper time}
d\tau=\frac{m A}{E^2-m^2A}ds.
\end{equation}

By introducing the inverse radial coordinate $u=1/r$, the orbit equation can be derived from Eqs.~\eqref{angular momentum} and~\eqref{radial equation} as follows:

\begin{eqnarray}
\label{trajectory equation}
\left(\frac{du}{d\varphi}\right)^2=C^2u^4\left[\left(\frac{\varepsilon}{h}\right)^2-A\left(\frac{1}{h^2}+\frac{1}{C}\right)\right].
\end{eqnarray}

Here, $h=J/m$ represents the angular momentum per unit mass, and $\varepsilon=E/m$ denotes the energy per unit mass. The energy and angular momentum for an asymptotic observer at infinity are given by~\cite{Crisnejo:2018uyn}:

\begin{eqnarray}
\label{enan}
&&E=\frac{m}{\sqrt{1-v^2}}~,~~J=\frac{m v b}{\sqrt{1-v^2}},
\end{eqnarray}

Here, $v$ represents the particle velocity, and $b$ is the impact parameter defined by:
\begin{eqnarray}
\label{impact}
\frac{J}{E}=vb.
\end{eqnarray}
via Eq.~\eqref{enan}, Jacobi metric~\eqref{SS-Jacobi-2} becomes
\begin{equation}
\label{Jacobi-metric}
ds^2=m^2\bigg(\frac{1}{1-v^2}-A\bigg)\bigg[\frac{1}{A^2}dr^2+\frac{C}{A}d\varphi^2 \bigg]~,
\end{equation}

\begin{widetext}
\begin{eqnarray}
\mathcal{K}&=& \frac{1}{3 v^{8} m^{4} r^{5}}4 (v^{4}-2 v^{2}+1) (((-M \,v^{4}+(14 M+r) v^{2}-16 M) \ln(r)+(2 M+\frac{r}{2}) v^{4}-(16 M+2 r) v^{2}+16 M) \zeta r \,q^{\frac{3}{2}} \notag \\&+&3 q^{2} (M+\frac{r}{4}) v^{4}+3 (-9 M \,q^{2}+\frac{1}{2} M \,r^{2}-q^{2} r) v^{2}+\frac{57 M \,q^{2}}{2})
\end{eqnarray}

\end{widetext}
and the trajectory equation~\eqref{trajectory equation} comes to
\begin{eqnarray}
\label{trajectory equation1}
\left(\frac{du}{d\varphi}\right)^2&=&C^2u^4\bigg[\frac{1}{v^2b^2}-A(R)\left(\frac{1-v^2}{v^2 b^2}+\frac{1}{C}\right)\bigg].
\end{eqnarray}

From Jacobi metric in Eq.~\eqref{Jacobi-metric}, one has
\begin{eqnarray}
\label{dds}
\frac{ds}{d\varphi}\bigg{|}_{C_R}=\left[m^2\bigg(\frac{1}{1-v^2}- A(R)\bigg)\frac{C(R)}{A(R)}\right]^{1/2}.
\end{eqnarray}

Furthermore, one can choose the velocity along the curve $C_R$ as $\dot{C}_{R}^i=\left(0,d\varphi (R)/ds\right)$, which satisfies unit speed condition $g_{ij}\dot{C}_{R}^{i}\dot{C}_{R}^{j}=1$.
This condition yields
\begin{eqnarray}
&&(\nabla_{\dot{C}_R}{\dot{C}_R})^{r}=\Gamma_{\varphi\varphi}^{r}(R)(\dot{C}_{R}^{\varphi})^2~,~~~(\nabla_{\dot{C}_R}{\dot{C}_R})^{\varphi}=0,
\end{eqnarray}
and now, the geodesic curvature of $C_R$ can be expressed as follows
\begin{eqnarray}
\label{geode-cur}
k_g(C_R)=\mid\nabla_{\dot{C}_R}{\dot{C}_R}\mid=\sqrt{g_{rr}\left(\Gamma_{\varphi\varphi}^{r}\right)^2}\left(\frac{d\varphi}{ds}\right)^2\bigg{|}_{C_R}.
\end{eqnarray}
Together with Eqs.~\eqref{dds} and~\eqref{geode-cur}, one can obtain
\begin{eqnarray}
&&\left({k_g\frac{ds}{d\varphi}}\right)\bigg{|}_{C_R}=\sqrt{\frac{1}{A(R) C(R)}\left(\Gamma_{\varphi\varphi}^{r}(R)\right)^2}~. \label{geodesic}
\end{eqnarray}

At this point, a special case can be considered
\begin{eqnarray}
\label{condition}
\lim_{R\rightarrow\infty}\left({k_g\frac{ds}{d\varphi}}\right)\bigg{|}_{C_R}=1.
\end{eqnarray}

This implies that the two-dimensional Jacobi geometry is asymptotically Euclidean. Subsequently, Eq.~\eqref{deflang} results in:

\begin{widetext}
\begin{eqnarray}
\label{GBT-K}
\alpha=-\lim_{R\rightarrow\infty}\iint_D{K}dS.
\end{eqnarray}

\begin{eqnarray}
& \alpha\simeq& \frac{9 q^{2} \pi}{4 b^{2} m^{4} v^{4}}-\frac{8 \ln(2) \zeta q^{\frac{3}{2}}}{3 v^{6} m^{4} b}+\frac{4 \zeta q^{\frac{3}{2}}}{3 v^{4} m^{4} b}+\frac{8 q^{\frac{3}{2}} \zeta \ln(b)}{3 b m^{4} v^{6}}-\frac{q^{2} \pi}{b^{2} m^{4} v^{6}}-\frac{16 q^{\frac{3}{2}} \zeta \ln(b)}{3 b m^{4} v^{4}}+\frac{16 \ln(2) \zeta q^{\frac{3}{2}}}{3 v^{4} m^{4} b}-\frac{16 M q^{\frac{3}{2}} \zeta \ln(2) \pi}{3 b^{2} m^{4} v^{8}}
\notag\\&-&
\frac{16 M q^{\frac{3}{2}} \pi \zeta \ln(b)}{3 b^{2} m^{4} v^{8}}+\frac{152 M q^{2}}{9 v^{8} m^{4} b^{3}}+\frac{16 M q^{\frac{3}{2}} \pi \zeta}{3 b^{2} m^{4} v^{8}}+\frac{4 M}{v^{6} m^{4} b}-\frac{8 M}{v^{4} m^{4} b}-\frac{496 q^{2} M}{9 v^{6} m^{4} b^{3}}-\frac{17 M \pi q^{\frac{3}{2}} \zeta \ln(b)}{b^{2} m^{4} v^{4}}\notag\\&-&\frac{17 M \pi q^{\frac{3}{2}} \zeta \ln(2)}{b^{2} m^{4} v^{4}}+\frac{49 M \pi q^{\frac{3}{2}} \zeta \ln(2)}{3 b^{2} m^{4} v^{6}}+\frac{49 M \pi q^{\frac{3}{2}} \zeta \ln(b)}{3 b^{2} m^{4} v^{6}}-\frac{18 M \pi q^{\frac{3}{2}} \zeta}{b^{2} m^{4} v^{6}}+\frac{188 q^{2} M}{3 v^{4} m^{4} b^{3}}+\frac{127 M \pi q^{\frac{3}{2}} \zeta}{6 b^{2} m^{4} v^{4}}
\end{eqnarray}
\end{widetext}

\begin{figure}
    \centering
\includegraphics[width=0.4\textwidth]{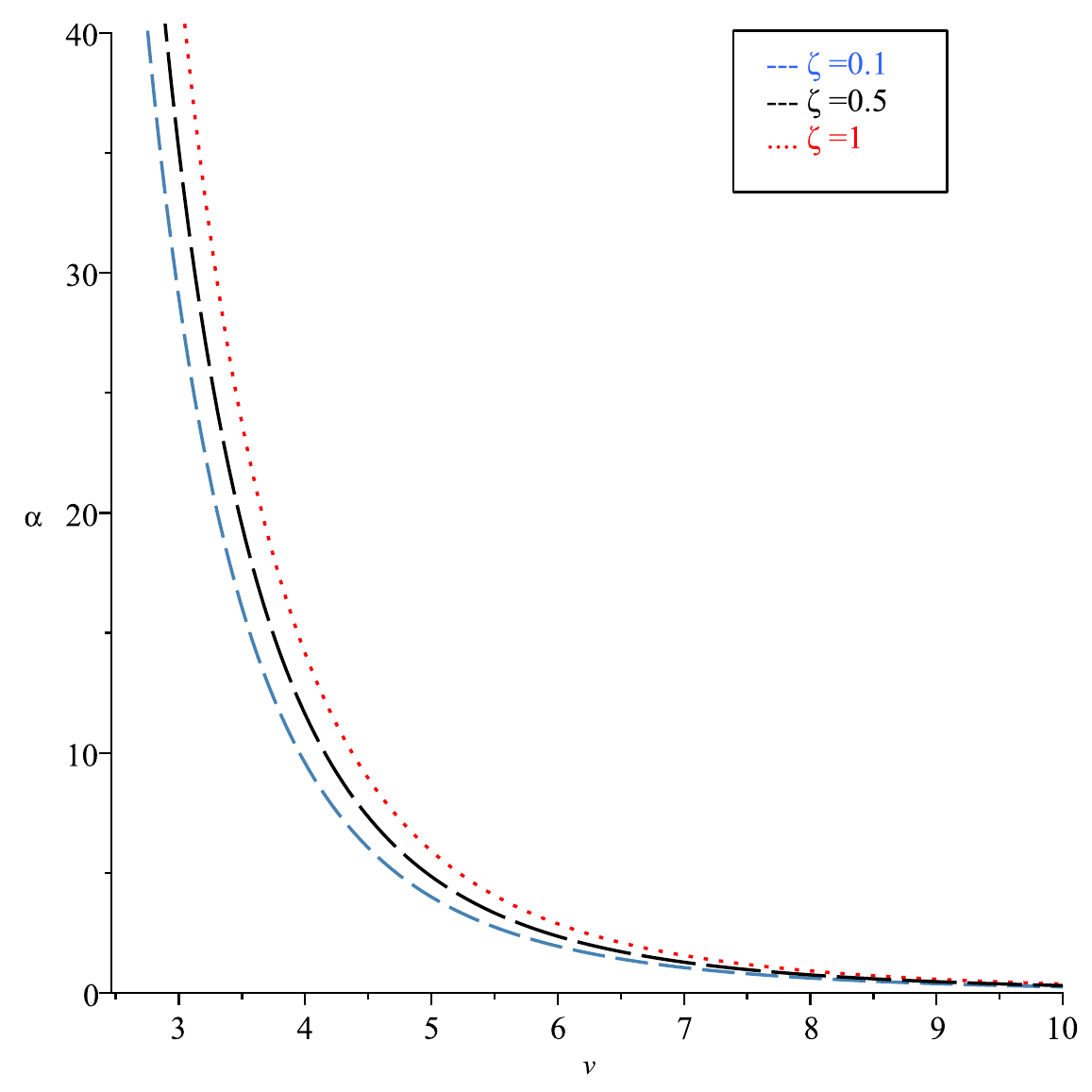}
    \caption{In the figure, the variation of the deflection angle versus velocity ($v$) is depicted by manipulating the Nonlinear Electrodynamics (NED) parameter $\zeta$, while maintaining fixed values for other parameters: $m=1$, $q=0.2$, $b=1$, and $M=1$. It is observed that for $\zeta>0$, the deflection angle $\alpha$ exhibits a gradual expansion.}
    \label{fig3}
\end{figure}

\begin{figure}
    \centering
    \includegraphics[width=0.4\textwidth]{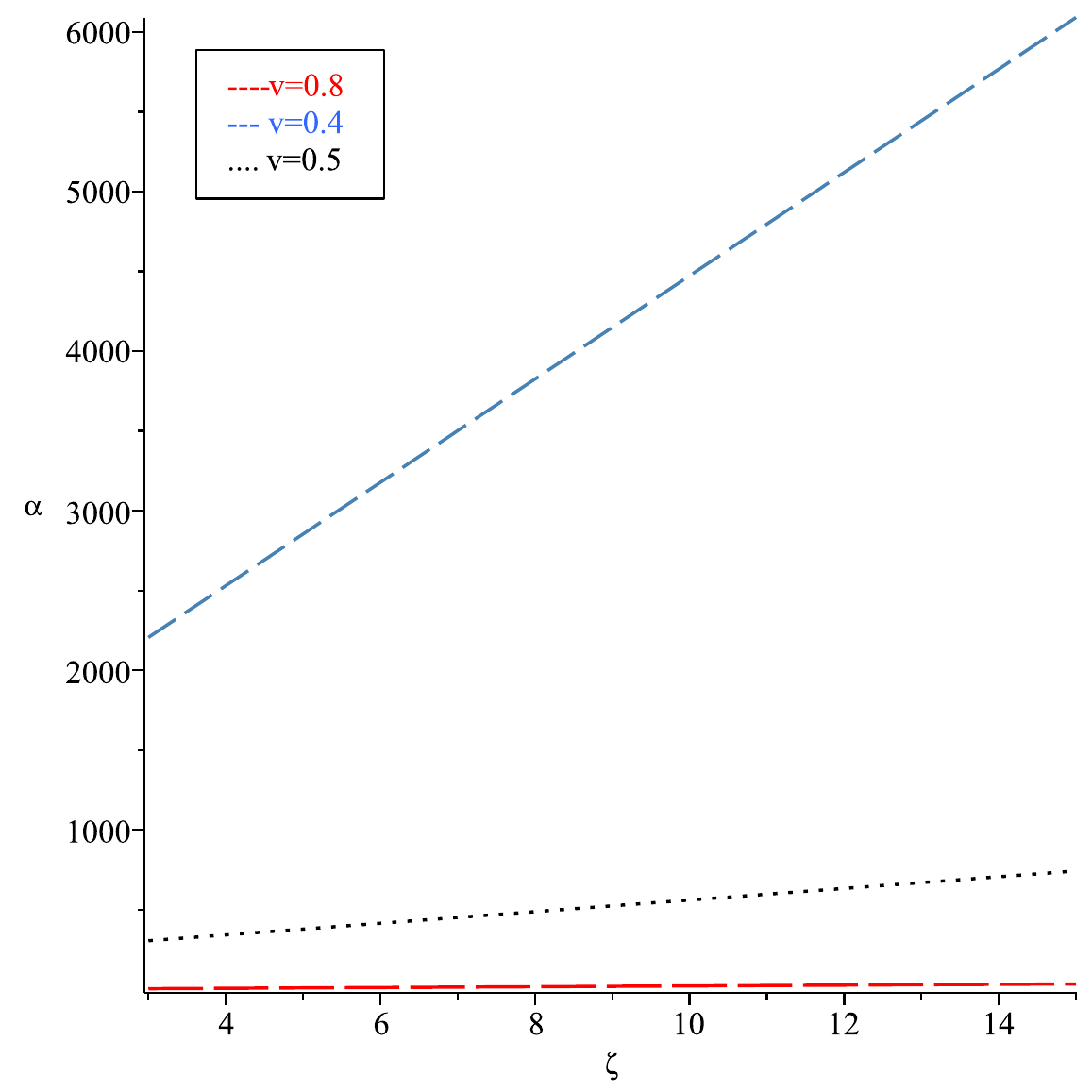}
    \caption{The graph illustrates the function $\alpha(\zeta)$ for various values of velocity $v$, with fixed parameters: $m=1$, $q=0.2$, $b=1$, and $M=1$.}
    \label{fig4}
\end{figure}

Our findings reveal a congruence between the derived expression and the outcome obtained through the application of the Gauss-Bonnet theorem to the optical metric~\cite{Gibbons:2008rj}. In accordance with Eq.~\eqref{GBT-K}, the computation of the deflection angle for massive particles involves integrating the intrinsic curvature of space. Notably, this integration spans an infinite Jacobi domain situated beyond the particle trajectory concerning the lens. Consequently, our results emphasize that the deflection angle can be interpreted as a distinctive global topological effect, as illuminated by Gibbons et al.~\cite{Gibbons:2008rj}. The behaviour of deflection angle with various parameters are shown in Fig.s \ref{fig3} and \ref{fig4}.

\section{Effect of the cold non-magnetized plasma medium on the deflection angle of NED black hole with quark-antiquark confinement}

In this section, we investigate the impact of a cold, non-magnetized plasma medium on the deflection angle of the Nonlinear Electrodynamics (NED) black hole with quark-antiquark confinement. \textcolor{black}{Note that gravitational lensing in plasma were considered earlier in \cite{Bisnovatyi-Kogan:2010flt}.} To initiate this study, we consider a cold non-magnetized plasma characterized by the refractive index $n$
\cite{Bisnovatyi-Kogan:2010flt,Crisnejo:2018uyn,Yasir:2023mdk}:

\begin{equation}
n^{2}(x, \omega(x))=1-\frac{\omega_{e}^{2}(x)}{\omega^{2}(x)}
\end{equation}
with
\begin{eqnarray}
\omega_{e}^{2}(x)=\frac{4 \pi e^{2}}{m_{e}} N(x)=K_{e} N(x).
\end{eqnarray}

Due to the gravitational redshift, the frequency of a photon at a specific radial position $r$ is expressed as:
\begin{equation}
\omega(r)=\frac{\omega_{\infty}}{\sqrt{A(r)}}
\end{equation}

Here, \(e\) and \(m_{e}\) denote the charge of the electron and its mass, respectively. \((\omega_{\infty})\) represents the photon frequency measured by an observer at infinity. This formulation implies that the refractive index \((n)\) solely exhibits radial dependence. \(N(x)\) stands for the number density of electrons in the plasma. It is noteworthy that only light rays with \(\omega(x)>\omega_{e}(x)\) can propagate through the plasma. Conversely, if \(\omega(x)<\omega_{e}(x)\), the refractive index becomes imaginary, rendering waves with such frequencies unable to propagate through the plasma and instead become evanescent.

Photons traveling through a plasma exhibit a deviation from null geodesics of the underlying space-time, and this deviation is dependent on the frequency of the photons. Additionally, even in the presence of a homogeneous plasma, characterized by $\omega_e=$ constant, if the underlying space-time induces a nontrivial gravitational redshift—meaning the photon frequency $\omega$ changes along the trajectory—this results in a nontrivial dispersion through Eq. (1), allowing a deviation of light rays from null geodesic trajectories. It is important to note that this particular effect is not present in a flat space-time. The refractive index for this black hole is given by\cite{Perlick:2015vta, Crisnejo:2018uyn}:

\begin{eqnarray}
n(r)=\sqrt{1-\frac{\omega_e^2}{\omega_{\infty}^2} A(r)}.
\end{eqnarray}

Here, $\omega_e$ represents the electron plasma frequency, and $\omega_{\infty}$ denotes the photon frequency measured by a static observer at infinity. The corresponding optical line element can be defined as:

\begin{eqnarray}
d\sigma^2&=&\gamma_{ij}dx^i dx^j=-\frac{n^2}{g_{00}}g_{ij}dx^i dx^j \\
&=&n^2\left(\frac{1}{A^2}dr^2+\frac{r^2}{A}d\phi^2\right).~\label{oppmetric}
\end{eqnarray}

This optical line element is conformally related to the induced metric on the spatial section with $\theta=\frac{\pi}{2}$.

\begin{widetext}

Subsequently, we compute the Gaussian curvature as:
\begin{eqnarray}
\tilde{\mathcal{K}}\!\!&=& \frac{\mathcal{R}_{r\phi r \phi}(g^{opt})}{det g^{opt}},
\end{eqnarray}

\begin{eqnarray}
\tilde{\mathcal{K}}&=& \frac{16 M \,q^{\frac{3}{2}} \ln(r) \zeta w_{e}^{2}}{r^{4} w_{\infty}^{2}}+\frac{4 M \,q^{\frac{3}{2}} \ln(r) \zeta}{r^{4}}-\frac{2 q^{\frac{3}{2}} \ln(r) \zeta w_{e}^{2}}{r^{3} w_{\infty}^{2}}-\frac{4 q^{\frac{3}{2}} \zeta \ln(r)}{3 r^{3}}-\frac{40 M \,q^{\frac{3}{2}} \zeta w_{e}^{2}}{3 r^{4} w_{\infty}^{2}}-\frac{8 M \,q^{\frac{3}{2}} \zeta}{3 r^{4}}\notag\\&+&\frac{10 q^{\frac{3}{2}} \zeta w_{e}^{2}}{3 r^{3} w_{\infty}^{2}}+\frac{2 q^{\frac{3}{2}} \zeta}{r^{3}}-\frac{26 M \,q^{2} w_{e}^{2}}{r^{5} w_{\infty}^{2}}-\frac{6 M \,q^{2}}{r^{5}}-\frac{3 M w_{e}^{2}}{r^{3} w_{\infty}^{2}}-\frac{2 M}{r^{3}}+\frac{5 q^{2} w_{e}^{2}}{r^{4} w_{\infty}^{2}}+\frac{3 q^{2}}{r^{4}}
\end{eqnarray}

in which 
\begin{equation}
   det(g^{opt})=\frac{n(r)^{4} r^{2}}{A(r)^{3}}
\end{equation}
\end{widetext}
Here, the plasma parameter is defined as $\delta\equiv \frac{\omega_e^2}{\omega_{\infty}^2}$. To ensure the propagation of a photon in the plasma, it is necessary to have $\omega_{\infty}\geq\omega_e$, leading to $0\leq\delta\leq1$. Additional details about the plasma can be found in Ref.~\cite{Bisnovatyi-Kogan:2010flt}. Moreover, from Eq.~(\ref{oppmetric}), it follows that:

\begin{eqnarray}
\frac{d\sigma}{d\phi}\bigg|_{\gamma_R}=n\sqrt{\frac{r^2}{A(r)}},
\end{eqnarray}
which results in
\begin{eqnarray}
\lim_{R\rightarrow \infty}\tilde{\kappa}(C_R) \frac{d\sigma}{d\phi}\bigg|_{\gamma_R}\approx 1.
\end{eqnarray}

By considering the zero-order particle trajectory $r=\frac{b}{\sin\phi}$ and taking the limit $R\rightarrow \infty$, the Gauss-Bonnet theorem can be expressed as:
\begin{eqnarray}
\int^{\pi+\alpha}_0 d\phi=\pi-\int^{\pi}_0\int^{\infty}_{\frac{b}{\sin\phi}}\tilde{\mathcal{K}}dS.
\end{eqnarray}

\begin{widetext}
Subsequently, the deflection angle can be calculated as:
\begin{eqnarray}
\alpha&=&-\int^{\pi}_0\int^{\infty}_{\frac{b}{\sin\phi}}\tilde{\mathcal{K}}dS \\
      &\simeq&\frac{4 q^{\frac{3}{2}} \zeta}{3 b}-\frac{3 \pi q^{2}}{4 b^{2}}-\frac{4 q^{\frac{3}{2}} \zeta \delta \ln(2)}{b}+\frac{4 q^{\frac{3}{2}} \zeta \delta \ln(b)}{b}-\frac{5 \pi q^{2} \delta}{4 b^{2}}-\frac{8 q^{\frac{3}{2}} \zeta \ln(2)}{3 b}+\frac{4 q^{\frac{3}{2}} \zeta \delta}{3 b}+\frac{8 q^{\frac{3}{2}} \zeta \ln(b)}{3 b}\notag\\&+&\frac{5 q^{\frac{3}{2}} \zeta M \delta \pi}{6 b^{2}}-\frac{5 q^{\frac{3}{2}} \zeta \ln(2) \delta \pi}{2 b^{2}}-\frac{5 q^{\frac{3}{2}} \zeta M \delta \pi \ln(b)}{2 b^{2}}+\frac{6 M \delta}{b}-\frac{4 M \,q^{2}}{3 b^{3}}-\frac{5 M \,q^{\frac{3}{2}} \pi \zeta}{6 b^{2}}+\frac{44 M \,q^{2} \delta}{9 b^{3}}+\frac{4 M}{b}.~\label{deflangp1}
\end{eqnarray}
\end{widetext}

\begin{figure}[!ht]
    \centering
\includegraphics[width=0.4\textwidth]{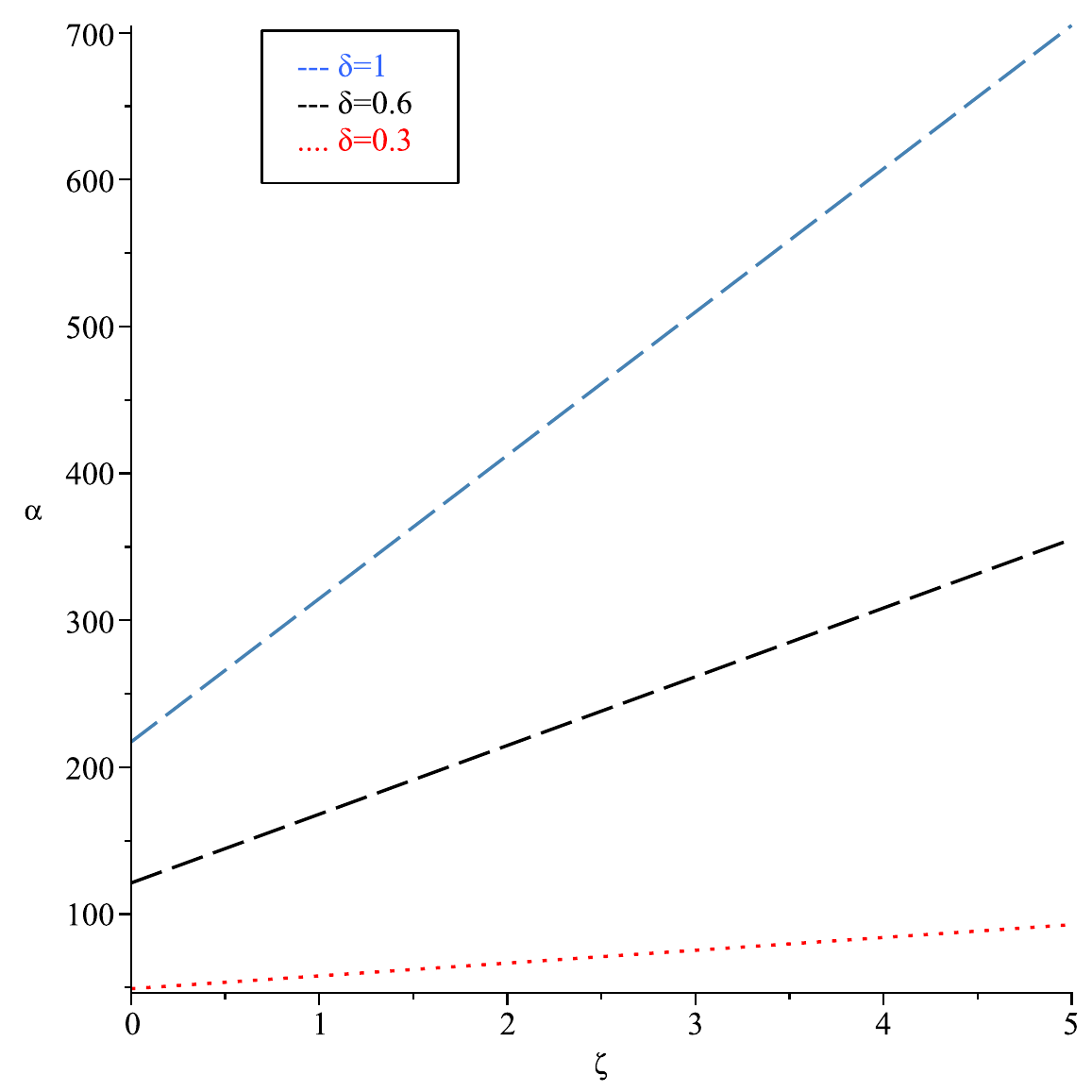}

    \caption{The figure illustrates the plots of the deflection angle versus impact parameter ($b$) for NED black holes in a cold non-magnetized plasma. The selected physical parameters include $M=1$, $b=0.1$, and $q=0.2$.}
    \label{fig5}
\end{figure}

It is straightforward to demonstrate that Eq.~(\ref{deflangp1}) reduces to Eq.~(\ref{deflang}) when $\delta\rightarrow 0$. Moreover, the deflection angle increases with the plasma parameter $\delta$, implying that for a fixed electron plasma frequency, the lower the photon frequency measured by a static observer at infinity, the larger the deflection angle will be shown in Fig. \ref{fig5}.

\section{Effect of the axion-plasmon medium on the deflection angle of NED black hole with quark-antiquark confinement}

Additionally, the exploration of axion-photon coupling is motivated by the string theory and the unkown of dark matters. Incorporating the axion-photon coupling into the electromagnetic framework introduces novel theoretical possibilities and phenomena. This generalization is inspired by works such as \cite{Mendonca:2019eke,Wilczek:1987mv}, emphasizing the importance of extending electromagnetic theories to account for the influence of axions. This motivation underscores the significance of exploring axion-photon coupling not only for its implications in dark matter research but also for its broader impact on our understanding of fundamental forces and interactions in the universe. Finally, our exploration encompasses a generalized electromagnetic theory that incorporates the axion-photon coupling, as proposed in works such as \cite{Mendonca:2019eke,Wilczek:1987mv}:
\begin{equation}
\mathcal{L}=R-\frac{1}{4}F_{\mu\nu}F^{\mu\nu}-A_\mu J_e^\mu+\mathcal{L}_\varphi+\mathcal{L}_{\text{int}}.
\end{equation}

In this context, the symbols $R$, $F_{\mu\nu}$, and $J_e^\mu$ represent the Ricci scalar, electromagnetic tensor, and the four-vector current of electrons, respectively. Simultaneously, the term $\mathcal{L}_\varphi$ is characterized by the axion Lagrangian density, denoted as $\nabla_\mu\varphi^*\nabla^\mu\varphi-m_\varphi^2|\varphi|^2$. Lastly, the interaction term $\mathcal{L}_{\text{int}}$ is defined as $-(g/4)\varepsilon^{\mu\nu\alpha\beta}F_{\alpha\beta}F_{\mu\nu}$, representing the photon-axion interaction, with $g$ representing the relevant coupling.

The Hamiltonian describing the motion of a photon orbiting a black hole enveloped by an axion-plasmon medium is expressed as per \cite{Synge:1960ueh}.
\begin{equation}
\mathcal H(x^\alpha, p_\alpha)=\frac{1}{2}\left[ g^{\alpha \beta} p_\alpha p_\beta - (n^2-1)( p_\beta u^\beta )^2 \right],
\label{generalHamiltonian}
\end{equation}
Here, $x^\alpha$ denotes the space-time coordinates, $p_\alpha$ and $u^\beta$ represent the four-momentum and four-velocity of the photon, respectively, and $n$ signifies the refractive index ($n=\omega/k$, where $k$ is the wave number). In the presence of an axion-plasmon contribution, the refractive index is articulated as per \cite{Mendonca:2019eke}:

\begin{eqnarray}
n^2&=&1- \frac{\omega_{\text{p}}^2}{\omega^2}-\frac{f_0}{\gamma_{0}}\frac{\omega_{\text{p}}^2}{(\omega-k u_0)^2}-\frac{\Omega^4}{\omega^2(\omega^2-\omega_{\varphi}^2)}\nonumber \\
&&-\frac{f_0}{\gamma_{0}}\frac{\Omega^4}{(\omega-k u_0)^2(\omega^2-\omega_{\varphi}^2)},
\label{eq:n1}
\end{eqnarray}
Expressed in relation to the plasma frequency $\omega^2_{p}(x^\alpha)=4 \pi e^2 N(x^\alpha)/m_e$ (where $e$ and $m_e$ denote the electron charge and mass, respectively, and $N$ represents the number density of electrons), the photon frequency $\omega(x^\alpha)$ is defined as $\omega^2=( p_\beta u^\beta )^2$, and the axion frequency is denoted as $\omega_{\varphi}^2$.%
The axion-plasmon coupling parameter is denoted as $\Omega=(gB_{0}\omega_{p})^{1/2}$, where $B_0$ represents the homogeneous magnetic field in the $z$-direction. Additionally, $f_0$ stands for the fraction of electrons in the beam propagating inside the plasma with velocity $u_0$, and $\gamma_0$ represents the corresponding Lorentz factor. Given the uncertainty regarding the role of the electron beam scenario near the black hole, we simplify the situation by setting $f_0=0$. Subsequently, we express (\ref{eq:n1}) as:

\begin{eqnarray}
n^2(r)&=&1- \frac{\omega_{\text{p}}^2(r)}{\omega(r)^2}-\frac{\Omega^4}{\omega(r)^2[\omega(r)^2-\omega_{\varphi}^2]},\nonumber \\
&&=1- \frac{\omega_{\text{p}}^2(r)}{\omega(r)^2}\left(1+\frac{ g^2B^2_0 }{\omega(r)^2-\omega_{\varphi}^2}\right)
,
\label{eq:n2}
\end{eqnarray}
with
\begin{equation}
\omega(r)=\frac{\omega_0}{\sqrt{A(r)}},\qquad  \omega_0=\text{const}.
\end{equation}
Experiments related to axion-plasmon conversion introduce the constraint on frequency scales $\omega_{\text{p}}^2\gg \Omega^2$ or $\omega_{\text{p}}\gg gB_0$ \cite{Mendonca:2019eke}.

The most straightforward model corresponds to a medium featuring an Axion-Plasmon composition. In this case, the refractive index is expressed as per \cite{Atamurotov:2021cgh}:

\begin{equation}
n(r)\simeq\sqrt{ 1-\frac{\omega_{\text{p}}^2}{\omega_0^2}A(r) \left(1+\frac{ \tilde{B}^2_0 }{1-\tilde{\omega}_{\varphi}^2}\right)}.\label{45}
\end{equation}

\begin{widetext}

We can reformulate the optical metric for the black hole surrounded by the plasma as:
\begin{eqnarray}\nonumber
dt^{2}&=&\left[1-\frac{\omega_{\text{p}}^2}{\omega_0^2}A(r) \left(1+\frac{ \tilde{B}^2_0 }{1-\tilde{\omega}_{\varphi}^2}\right)\right]\Big[ \frac{dr^2}{A(r)^2} +\frac{r^2}{A(r)}d\phi^2\Big].
\end{eqnarray}

\begin{eqnarray}
\tilde{\mathcal{K}}&=&\frac{2 \left(\left(\left(B_{0}^{2}-\omega_{\phi}^{2}+1\right) \omega_{p}^{2}-\frac{2 \omega_{0}^{2} \omega_{\phi}^{2}}{3}+\frac{2 \omega_{0}^{2}}{3}\right) \ln \! \left(r \right)+\left(-\frac{5 B_{0}^{2}}{3}+\frac{5 \omega_{\phi}^{2}}{3}-\frac{5}{3}\right) \omega_{p}^{2}+\omega_{0}^{2} \omega_{\phi}^{2}-\omega_{0}^{2}\right) \zeta q^{\frac{3}{2}}}{r^{3} \omega_{0}^{2} \left(\omega_{\phi}^{2}-1\right)}
\end{eqnarray}

Then the deflection angle can be obtained as
\begin{eqnarray}
\alpha=-\int^{\pi}_0\int^{\infty}_{\frac{b}{\sin\phi}}\tilde{\mathcal{K}}dS 
 \simeq -\frac{4 q^{\frac{3}{2}} \zeta B_{0}^{2} \omega_{p}^{2} \ln\! \left(2\right)}{b \omega_{0}^{2}}+\frac{4 q^{\frac{3}{2}} \zeta \ln\! \left(b\right) B_{0}^{2} \omega_{p}^{2}}{b \omega_{0}^{2}}+\frac{4 q^{\frac{3}{2}} \zeta B_{0}^{2} \omega_{p}^{2}}{3 b \omega_{0}^{2}}-\frac{8 q^{\frac{3}{2}} \zeta \ln\! \left(2\right)}{3 b}-\frac{4 q^{\frac{3}{2}} \zeta \omega_{p}^{2} \ln\! \left(2\right)}{b \omega_{0}^{2}}\notag\\+\frac{8 q^{\frac{3}{2}} \zeta \ln\! \left(b\right)}{3 b}+\frac{4 q^{\frac{3}{2}} \zeta \ln\! \left(b\right) \omega_{p}^{2}}{b \omega_{0}^{2}}+\frac{4 q^{\frac{3}{2}} \zeta}{3 b}+\frac{4 q^{\frac{3}{2}} \zeta \omega_{p}^{2}}{3 b \omega_{0}^{2}}+\frac{3 \pi  q^{\frac{3}{2}} \zeta M \ln\! \left(2\right) B_{0}^{2} \omega_{p}^{2}}{2 b^{2} \omega_{0}^{2}}+\frac{3 \pi  q^{\frac{3}{2}} \zeta M \ln\! \left(b\right) B_{0}^{2} \omega_{p}^{2}}{2 b^{2} \omega_{0}^{2}}\notag\\-\frac{5 \pi  q^{\frac{3}{2}} \zeta M B_{0}^{2} \omega_{p}^{2}}{2 b^{2} \omega_{0}^{2}}+\frac{\pi  q^{\frac{3}{2}} \zeta M \ln\! \left(2\right)}{b^{2}}+\frac{3 \pi  q^{\frac{3}{2}} \zeta M \ln\! \left(2\right) \omega_{p}^{2}}{2 b^{2} \omega_{0}^{2}}+\frac{\pi  q^{\frac{3}{2}} \zeta M \ln\! \left(b\right)}{b^{2}}+\frac{3 \pi  q^{\frac{3}{2}} \zeta M \ln\! \left(b\right) \omega_{p}^{2}}{2 b^{2} \omega_{0}^{2}}\notag\\-\frac{3 \pi  q^{\frac{3}{2}} \zeta M}{2 b^{2}}-\frac{5 \pi  q^{\frac{3}{2}} \zeta M \omega_{p}^{2}}{2 b^{2} \omega_{0}^{2}}+\mathcal{O}(M^2,q^2).~\label{deflangp}
\end{eqnarray}
\end{widetext}

\begin{figure}
    \centering
\includegraphics[width=0.4\textwidth]{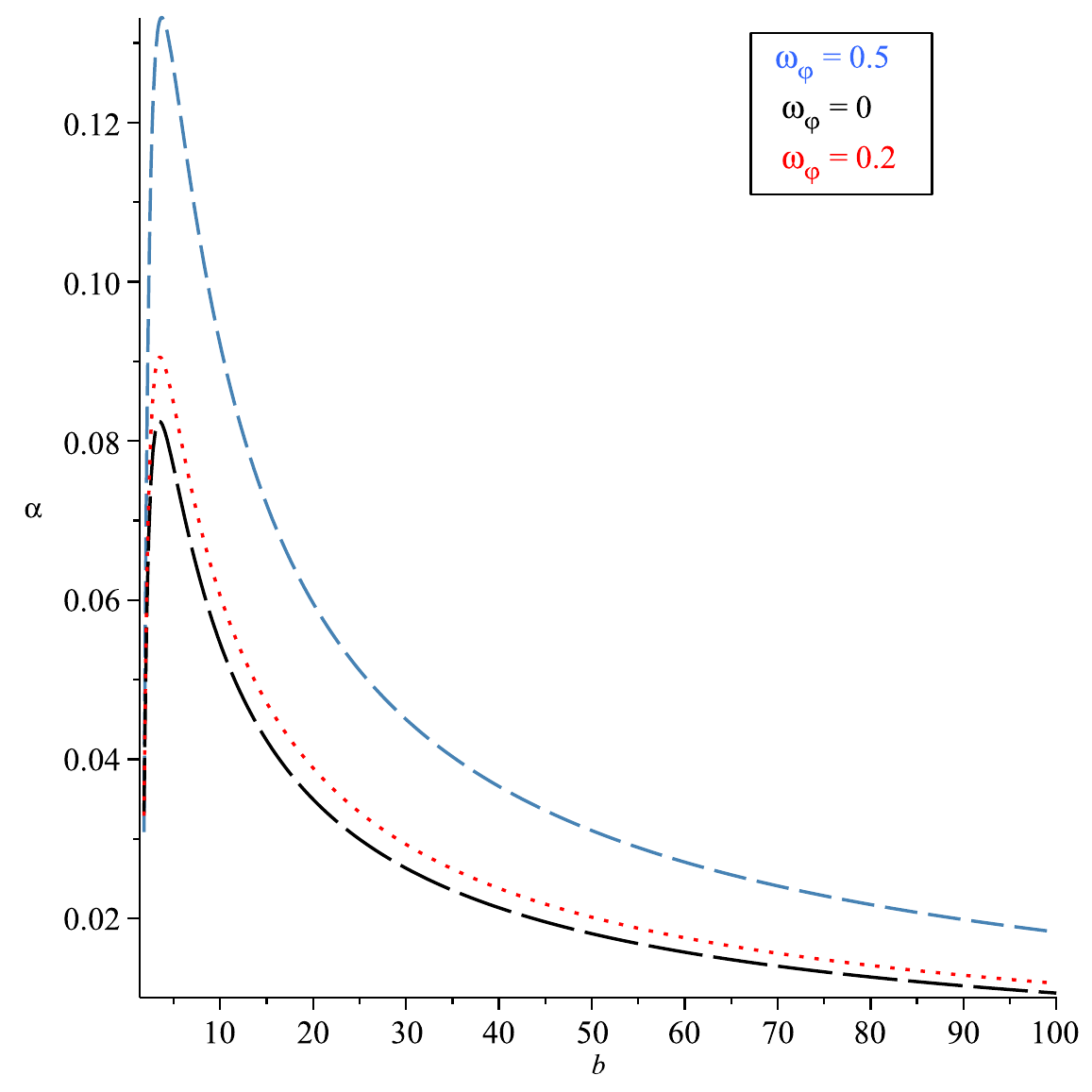}
    \caption{Figure illustrating the relationship between the deflection angle ($\alpha$) and the impact parameter ($b$) for various constant values of $\omega _ {\varphi}$.}
    \label{fig6}
\end{figure}

\begin{figure}
    \centering
\includegraphics[width=0.4\textwidth]{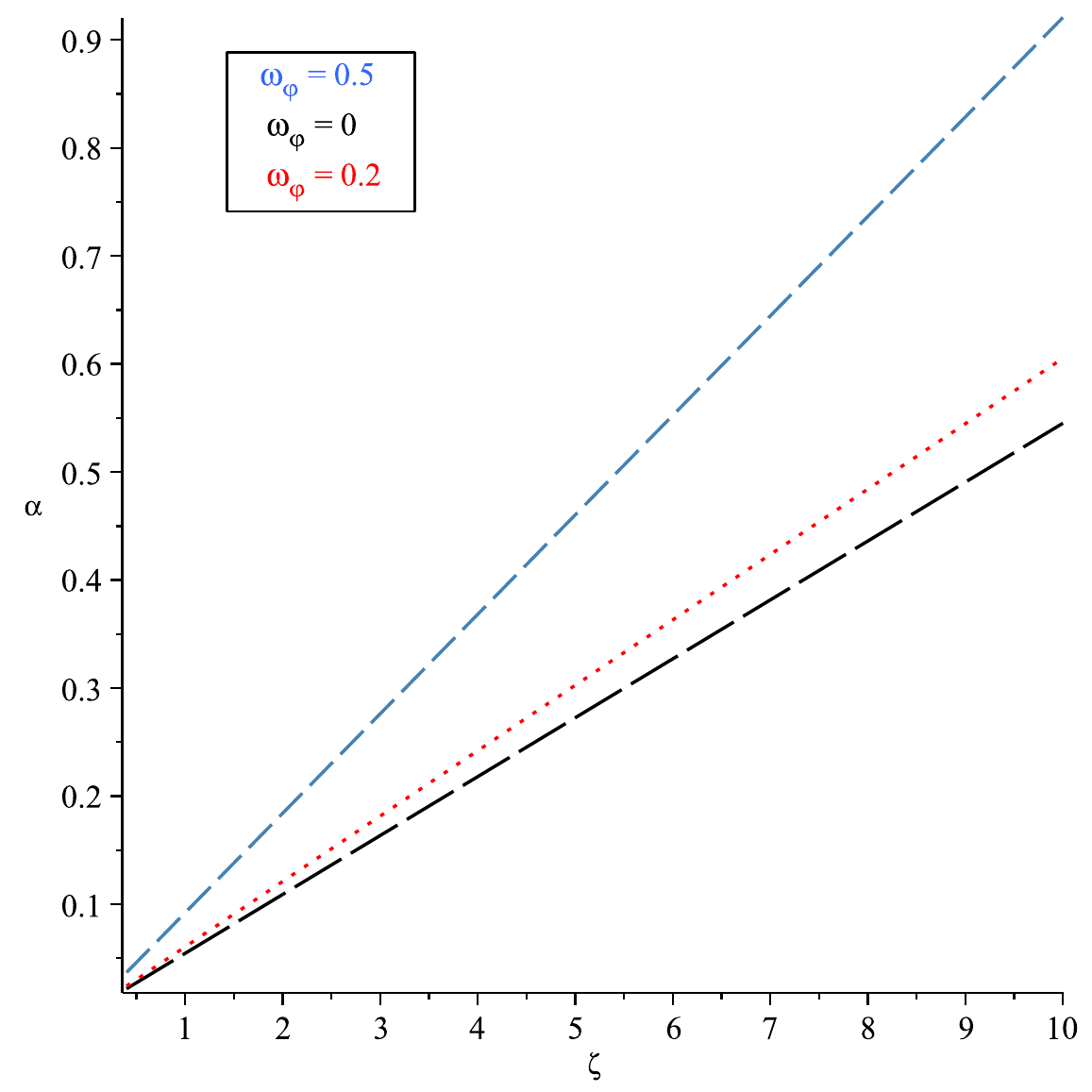}
    \caption{Plot depicting the deflection angle ($\alpha$) in relation to the NED parameter ($\zeta$), with varying constants such as $M=1$, $m=1$, $b=10$, $B_{0}=1$, $q=0.2$, and $\omega _ {0} =1$.}
    \label{fig7}
\end{figure}

\begin{figure}
    \centering
    \includegraphics[width=0.4\textwidth]{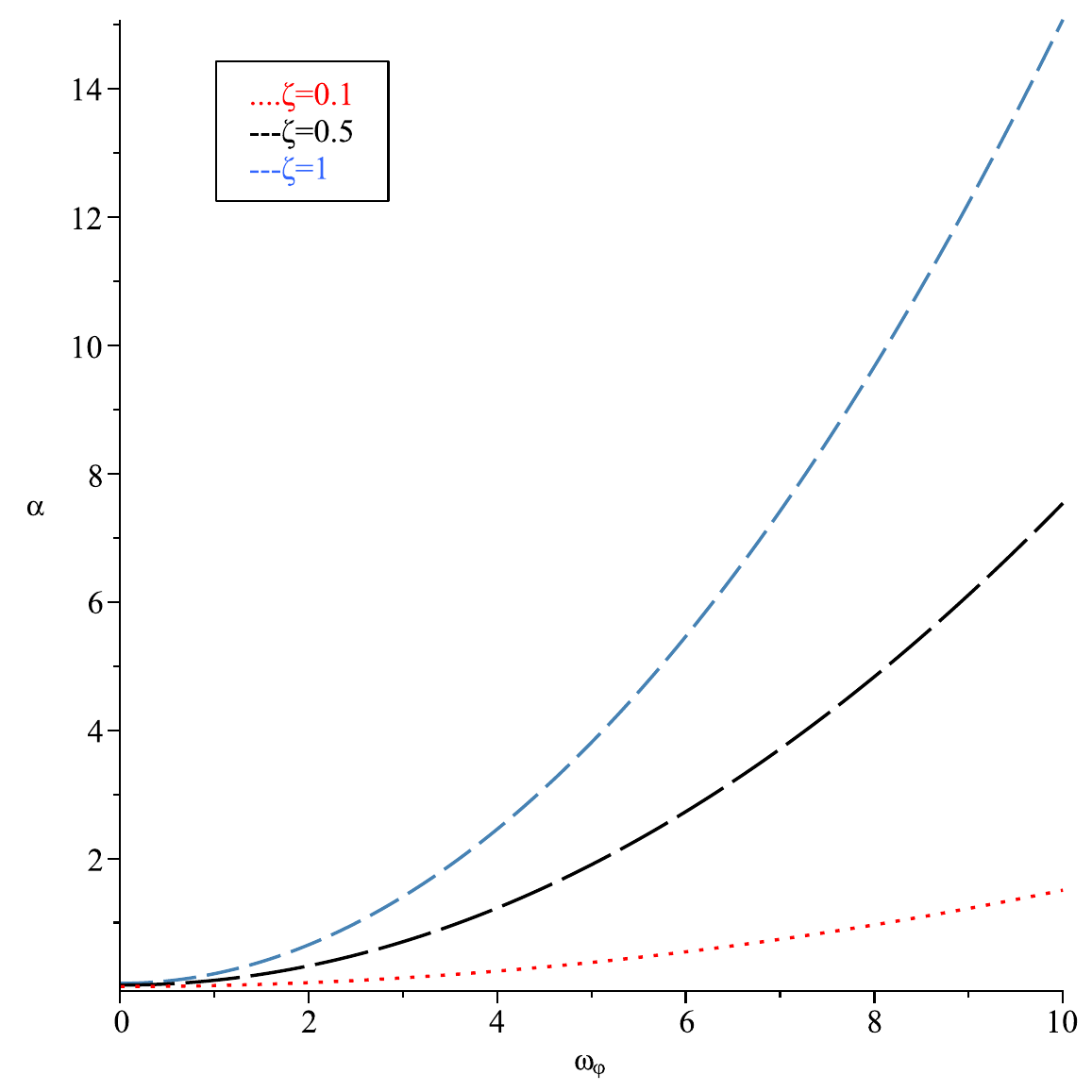}
    \caption{Plotting the deflection angle against $\omega_{\varphi}$ reveals that for fixed NED parameter values ($\zeta$ $=$ 0.1, 0.5, and 1), the deflection angle for the NED black hole exhibits an increase with the escalating magnitude of $\omega_{\varphi}$.}
    \label{fig8}
\end{figure}

In the Fig.s \ref{fig6}, \ref{fig7} and \ref{fig8}, we show the behaviour of deflection angle in the axion-plasmon medium.

\section{Conclusion} \label{sec8}

In this study, we conducted a comprehensive analysis of the impact of quark-antiquark confinement on the deflection angle exhibited by a Nonlinear Electrodynamics (NED) black hole. Initially, we explored the characteristics of the NED black hole incorporating quark-antiquark confinement. Subsequently, we introduced a formula (\ref{temperature}) derived from scrutinizing the Euler characteristic of the black hole spacetime. This formula provides a straightforward approach for computing Hawking temperatures in various coordinate systems.

In the subsequent section, we employed the Gauss-Bonnet theorem to calculate the deflection angle for the NED black hole. Notably, we observed that, for a constant value of $\zeta$, the deflection angle decreases with an increasing magnitude of $q$ and the impact parameter $b$.

We established the associated orbit equation in the equatorial plane and derived the static and spherically symmetric Jacobi metric. Utilizing Jacobi geometry, we formulated the expression for determining the deflection angle from the Gaussian curvature. Our findings indicated that the deflection angle $\alpha$ monotonically increases with the rising magnitude of $\zeta$ for a fixed value of $v$, and as $v$ values increased, the rate of increase in $\alpha$ decelerated. Exploring the influence of a cold non-magnetized plasma medium on gravitational lensing, we examined the deflection angle as defined in equation (\ref{deflangp1}). It was evident that equation (\ref{deflangp1}) converged to equation (\ref{deflang}) as $\delta$ approached zero, underscoring the vanishing of the plasma effect when $\delta=\frac{\omega_{e}}{\omega_{\infty}}\rightarrow 0$.

In conclusion, our investigation into the axion-plasmon effect on the optical properties of the NED black hole, particularly its influence on the gravitational lensing, holds significance in the context of the axion's role as a dark matter candidate. The potential to detect axion effects through gravitational lensing, particularly if it is coupled with photons, underscores the importance of further research in this direction. The insights gained from our analysis, along with the implications for future studies, highlight the intricate interplay between axions, plasmons, and gravitational lensing, providing a rich avenue for continued exploration in both theoretical and observational realms. Future research endeavors may delve deeper into these phenomena, potentially uncovering novel aspects of dark matter and contributing to our broader understanding of fundamental astrophysical processes.

\acknowledgments 
A. {\"O}. would like to acknowledge the contribution of the COST Action CA21106 - COSMIC WISPers in the Dark Universe: Theory, astrophysics and experiments (CosmicWISPers) and the COST Action CA22113 - Fundamental challenges in theoretical physics (THEORY-CHALLENGES). We also thank TUBITAK and SCOAP3 for their support.

\bibliography{ref}
\end{document}